\documentclass[onecollarge,natbib]{svjour2}
\bibpunct{[}{]}{;}{n}{}{,} 
\smartqed  
\usepackage{graphicx}
\usepackage{comment}
\usepackage{hyperref}
\usepackage{color}
\usepackage{subfigure}

\newcommand{\F}{F_{\pi\rho}}

\newcommand{\Epipi}{E_{\pi\pi}}

\usepackage{amssymb,amsmath}

\journalname{Archive of Applied Mechanics}
\begin{document}

\title{Meson electro-/photo-production from QCD}


\author{Ra\'ul A. Brice\~no\\
}

\institute{
Thomas Jefferson National Accelerator Facility,\\ 12000 Jefferson Avenue, Newport News, VA 23606, USA
\\
Department of Physics, Old Dominion University, Norfolk, VA 23529, USA
\\
\email{rbriceno@jlab.org}
}

\date{\color{white} Received:  / Accepted: }
 
\maketitle

\begin{abstract}
I present the calculation of the $\pi^+\gamma^\star\to\pi^+\pi^0$ transition amplitude from quantum chromodynamics performed by the Hadron Spectrum Collaboration. The amplitude is determined for a range of values of the photon virtuality and the final state energy. One observes a clear dynamical enhancement due to the presence of the $\rho$ resonance. By fitting the transition amplitude and analytically continuing it onto the $\rho$-pole, the $\rho\to\pi\gamma^\star$ form factor is obtained. This exploratory calculation, performed using lattice quantum chromodynamics, constitutes the very first determination of an electroweak decay of a hadronic resonance directly from the fundamental theory of quarks and gluons. In this talk, I highlight some of the necessary steps that made this calculation possible, placing emphasis on recently developed formalism. Finally, I discuss the status and outlook of the field for the study of $N\gamma^\star\to N^\star\to N\pi$ transitions.

\end{abstract}

\section{Introduction}
\label{intro}

The study of hadronic resonances is entering an exciting era. For the first time since the identification of quantum chromodynamics (QCD) as the fundamental theory of the strong interaction, one can hope to study hadronic resonances and their properties in a systematically controlled fashion. This is in part due to the tremendous progress made by the lattice QCD community.

The need for rigorous determinations of resonant properties directly from the standard model expands a wide range of phenomenology. These include the field of hadron spectroscopy (e.g., exotic hadrons~\cite{Briceno:2015rlt, Swanson:2015wgq, Prelovsek:2013cra}), hadronic structure (e.g., $N\to N^\star$ transitions) and heavy meson decays (e.g., $B\to K^\star\ell^+\ell^-$ weak decays~\cite{Horgan:2013pva}), among others. The theoretical progress towards the study of these processes has been historically limited by the fact that QCD is non-perturbative at low to medium energies. Presently, there is only one tool at our disposal that allows for the reliable study of QCD in this kinematic regime, this is the aforementioned lattice QCD.

By definition, lattice QCD requires one to place the theory in a finite, discretized Euclidean spacetime. Discretizing the spacetime introduces a UV cutoff, normally referred to as the lattice spacing. The truncation of the spacetime introduces an IR cutoff. For concreteness, I will only consider cubic volumes in the spacial extent with length $L$ and a temporal extent of length $T$. 

\begin{figure}[t]
\centering
\subfigure[]{\includegraphics[scale=0.35]{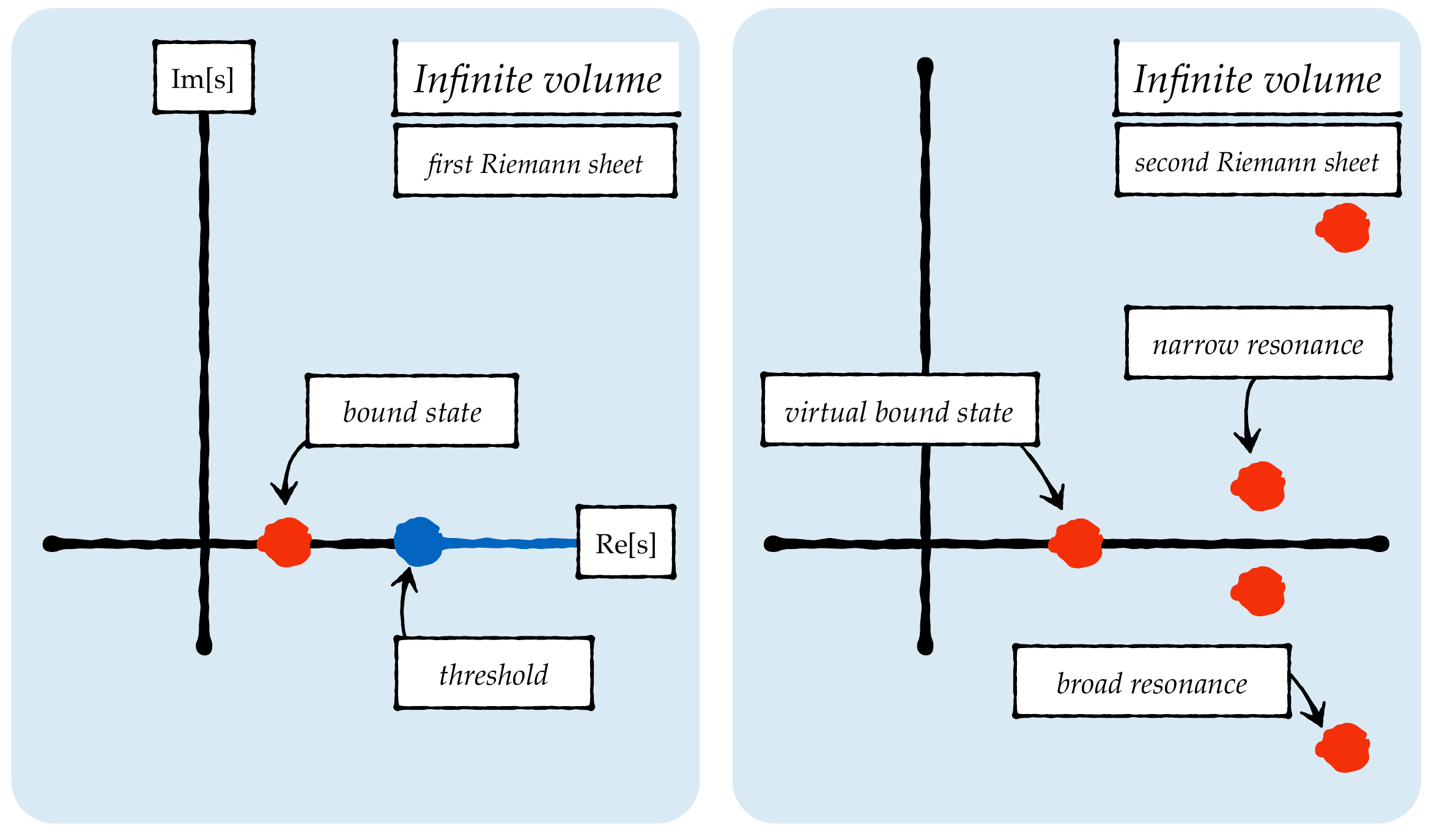}
\label{fig:IV_spec}}
\hspace{1cm} 
\subfigure[]{\includegraphics[scale=0.35]{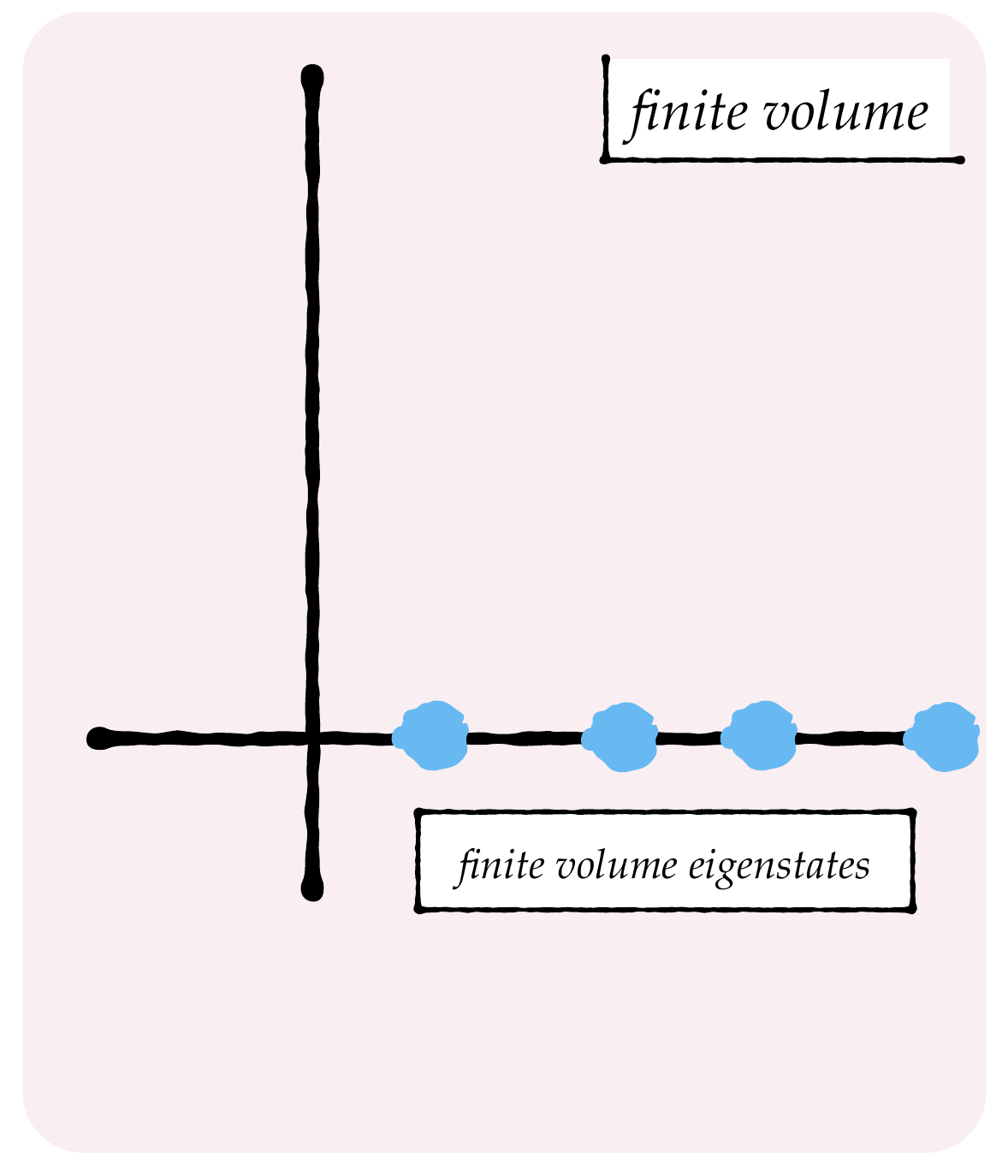}
\label{fig:FV_spec}}
\caption{Shown is the analytic structure of the spectrum of a generic quantum field theory in (a) the infinite-volume limit and (b) a finite volume. On the left, the red dots are meant to poles, while the blue dots denotes a branch-points associated with the opening of threshold. Along the real axis above the thresholds, there is a continuum of states. Poles on the real axis are bound states, while poles with non-zero imaginary components are resonances, these necessarily must lie on the second Riemann sheet.  In a finite volume, there is a discrete number of states and they lie on the real axis. In the text I describe the interpretation of finite volume states that lie above two-body thresholds. }
\label{fig:toy_spec} 
\end{figure}

The fact that the volume is finite leads to a drastic alteration of the theory's analytic structure. To illustrate this, it is sufficient to consider a generic quantum field theory. If we assume a theory that in the infinite volume has a bound state, followed by multiple thresholds, narrow and broad resonances, its spectrum can be qualitatively depicted by Fig.~\ref{fig:IV_spec}. The bound state would appear as an S-matrix pole on the real axis below all open thresholds, the thresholds emerge as branch-points whose cuts are commonly aligned along the positive Re[s]-axis, and resonances appear as poles off the Re[s]-axis. For resonances that lie above just one open thresholds, there will be two poles associated with the resonances appearing in the second Riemann sheet. These two poles correspond to complex conjugates of each other. Poles associated with a narrow resonance will be comparatively close to the real axis than the broader resonances. Experimentally, one only has access to quantities along the Re[s]-axis above thresholds, and resonant poles must deduced by analytically continuing fits of the S-matrix. 

There are two important consequences of placing a theory in a finite volume. The first is that, as is illustrated in Fig.~\ref{fig:FV_spec}, the spectrum becomes discretized and all poles reside on the Re[s]-axis. This emphasizes the fact that \emph{there are no hadronic resonances in a finite volume}. Second, in a finite volume there is no means of defining asymptotic states, and one cannot ``\emph{perform scattering}" in a finite volume. 

The only means to circumvent this limitation is to find non-perturbative relations between finite-volume quantities and infinite-volume observables. For the spectrum, this is typically attributed to L\"uscher in the literature~\cite{Luscher:1986pf, Luscher:1990ux}, and as a result I will refer to this formalism and its extensions~\cite{Rummukainen:1995vs, Kim:2005gf, Christ:2005gi, Briceno:2012yi, Hansen:2012tf} as \emph{L\"uscher formalism}. The most general extension of this formalism was presented in Ref.~\cite{Briceno:2014oea}, which accommodates for any number of two-body coupled channels. For electromagnetic transitions in the presence of an external current one needs another formalism to related finite-volume matrix elements of the current to infinite-volume amplitudes. This was first addressed by Lellouch and L\"uscher~\cite{Lellouch:2000pv} for $K\to\pi\pi$ weak decays.  This formalism has been since extended to accommodate increasingly complex systems~\cite{Lin:2001ek, Kim:2005gf, Christ:2005gi, Meyer:2012wk, Hansen:2012tf}. The most general formalism for $1\to2$~\cite{Briceno:2015csa, Briceno:2014uqa} and $2\to 2$ elastic/inelastic reactions~\cite{Briceno:2015tza} has been recently derived.~\footnote{Reference~\cite{Briceno:2015csa} also presented the most general result for $0\to2$ transitions, which was first presented in Ref.~\cite{Meyer:2011um} and later implemented in the study of $\gamma^\star\to\pi\pi$ in Ref.~\cite{Feng:2014gba, Bulava:2015qjz}.}

In Sec.~\ref{sec:formalism}, I give some details about the necessary formalism for the analysis of the spectrum and matrix elements, and in Sec.~\ref{sec:pigamma_to_pipi} I review its implementation in the calculation of the $\pi^+\gamma^\star\to\pi^+\pi^0$ transition amplitude, presented in Ref.~\cite{Briceno:2015dca, Briceno:2016kkp} by the Hadron Spectrum Collaboration. Finally, in Sec.~\ref{sec:outlook} I give a biased outlook for similar calculations of $N\to N^\star$ transitions from lattice QCD.

\section{Finite-volume formalism \label{sec:formalism}}

\begin{figure}[t]
\centering
\subfigure[]{\includegraphics[scale=0.25]{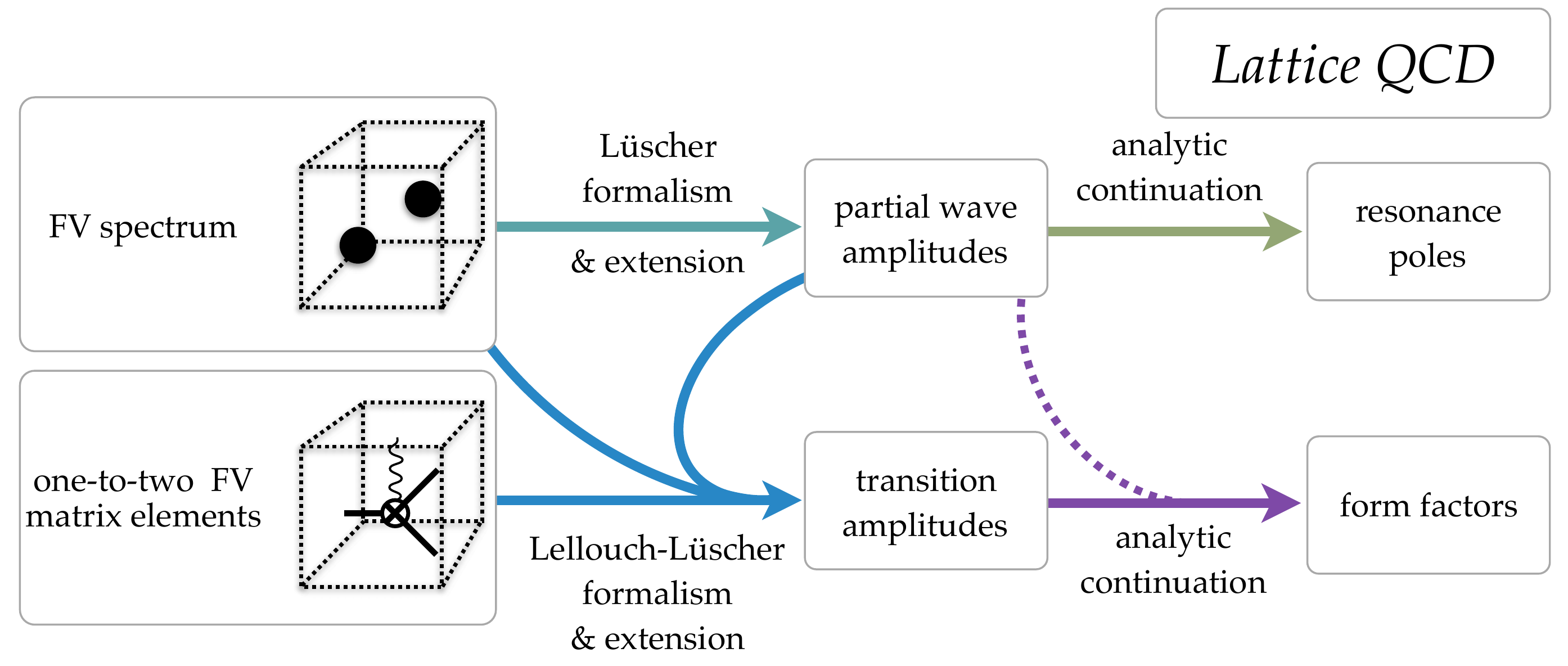}
\label{fig:LQCD_flow}}
\hspace{.5cm} 
\subfigure[]{\includegraphics[scale=0.25]{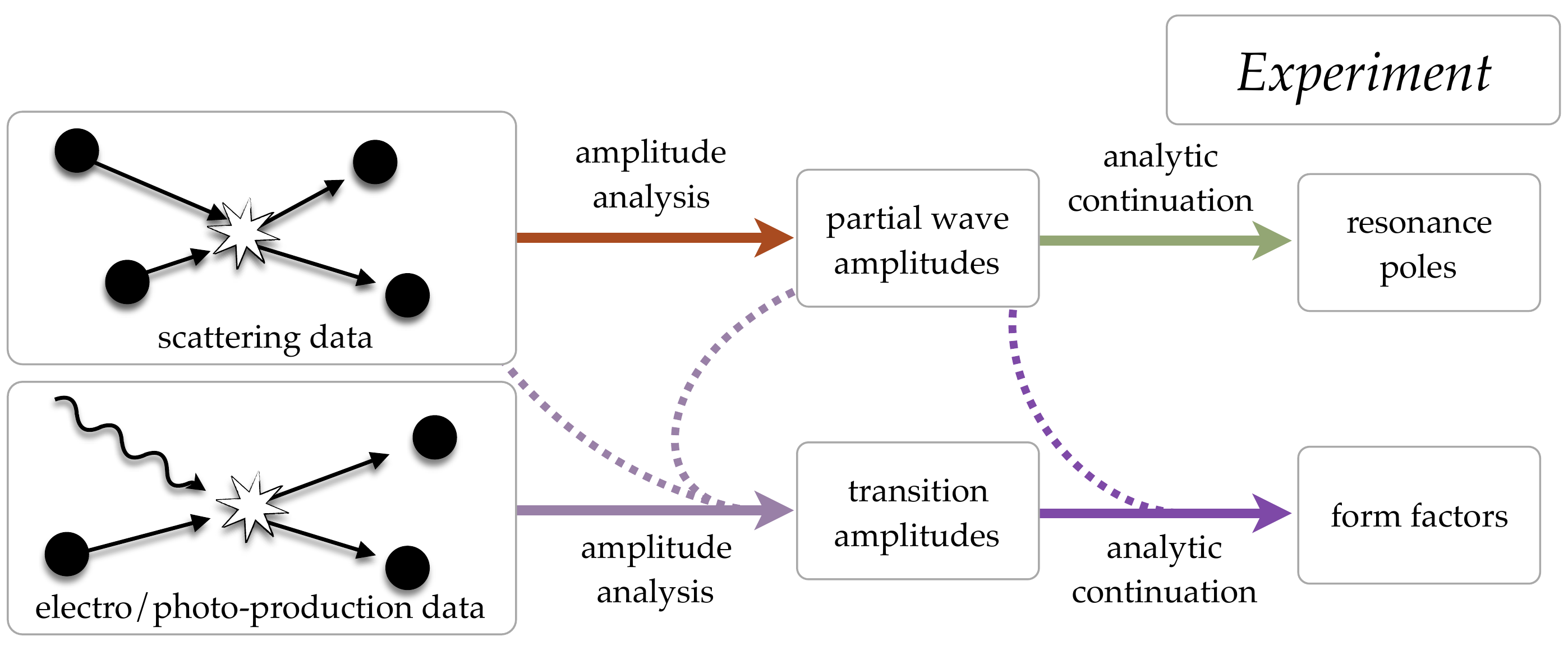}
\label{fig:Exp_flow}}
\caption{ (a) Shown is a flow chart depicting the procedure for obtaining resonance pole and form factors from lattice QCD. This is compared to the procedure for obtaining these same observables from experiment (b). Each step is further explained in the text.  }
\label{fig:flow_charts} 
\end{figure}

Here I give a \emph{bird's-eye view} of the steps needed at arriving at a resonant amplitude or a form factor of a resonance from lattice QCD.
\footnote{For more detailed discussions on the topic, I point the reader to recent reviews on the topic~\cite{Hansen:2015azg, Briceno:2014pka, Briceno:2014tqa, Prelovsek:2014zga, Mohler:2012nh}} 
To supplement the discussion, it is useful to look at Fig.~\ref{fig:flow_charts}, where I give a schematic flowchart for the steps needs to go from lattice QCD quantities to physical observables. For comparison, I also present a similar flowchart for experiments. I will refer to these figures periodically through the text.

I begin by reviewing the necessary formalism to obtain $2\to2$ scattering amplitudes from lattice QCD calculations. One can show that a finite-volume two-particle energy, $E_L$, satisfies the following relation~\cite{Luscher:1986pf, Luscher:1990ux, Rummukainen:1995vs, Kim:2005gf, Christ:2005gi, Briceno:2012yi, Hansen:2012tf, Briceno:2014oea}
\begin{equation}
\label{eq:QC_master}
\det[F^{-1}(E_L,L) + \mathcal M(E_L)] = 0\,,
\end{equation}
where $\mathcal M$ is the infinite-volume scattering amplitude, $F(E_L,L)$ is a known function of the volume, and the determinant acts on the space of partial waves and open channels, of which can be an infinite number of both. In general, due to the reduction of rotational symmetry, different partial waves mix. 

This equation qualitatively can be understood as follows. If one has access to the spectrum, one can constrain the scattering amplitude.  Equivalently one can constrain the spectrum if one has the scattering amplitude. This is what is being depicted by the light green arrow on the upper panel of Fig.~\ref{fig:LQCD_flow}. Obtaining partial wave amplitude from the lattice QCD finite-volume spectrum is analogous to the determination of partial wave amplitudes from experimental scattering data (depicted by the red arrow of the upper panel of Fig.~\ref{fig:Exp_flow}).

\begin{figure}
\centering
\subfigure[]{\includegraphics[scale=0.42]{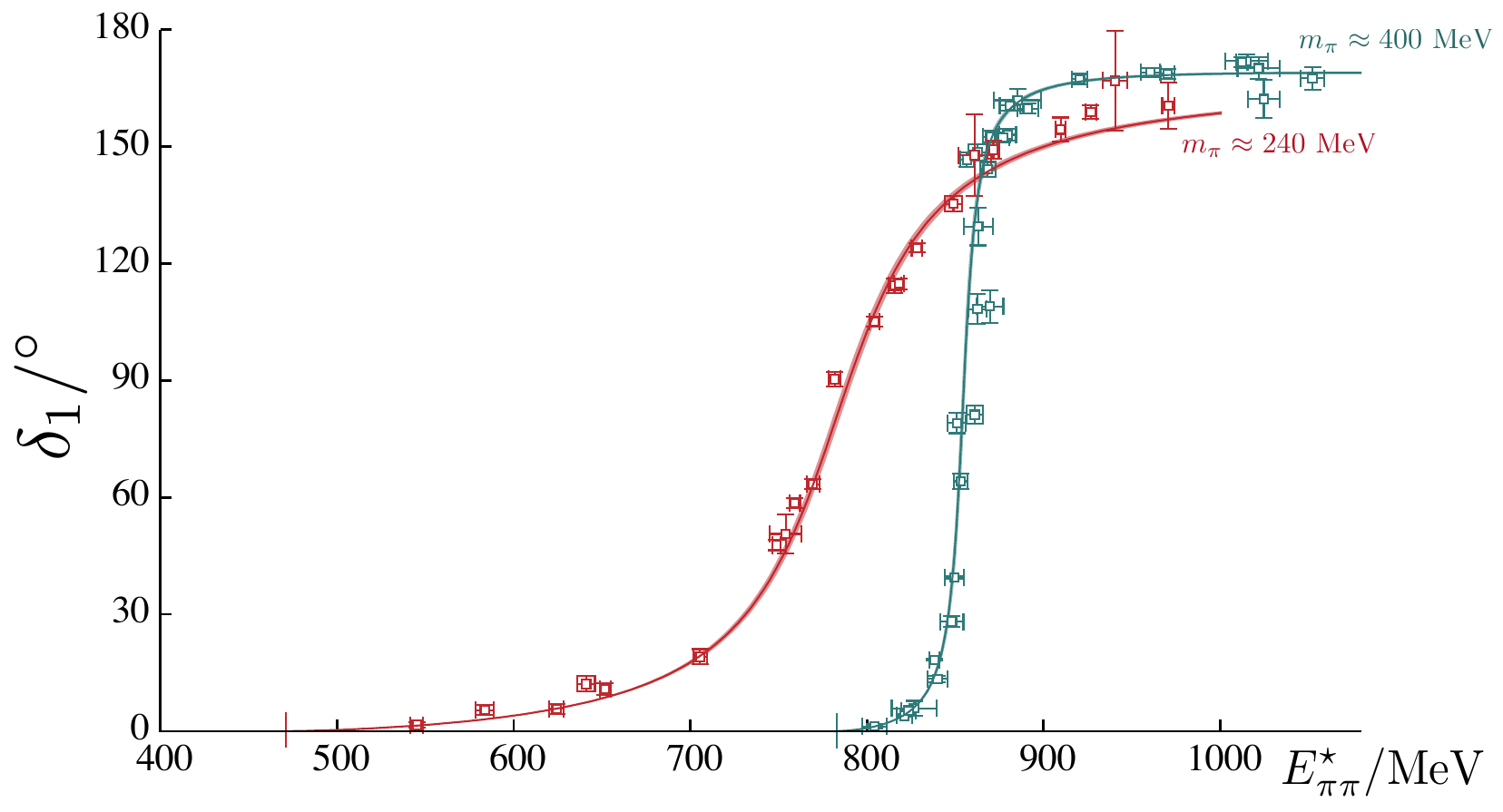}
\label{fig:860_840_scale_set}}
\subfigure[]{\includegraphics[scale=0.42]{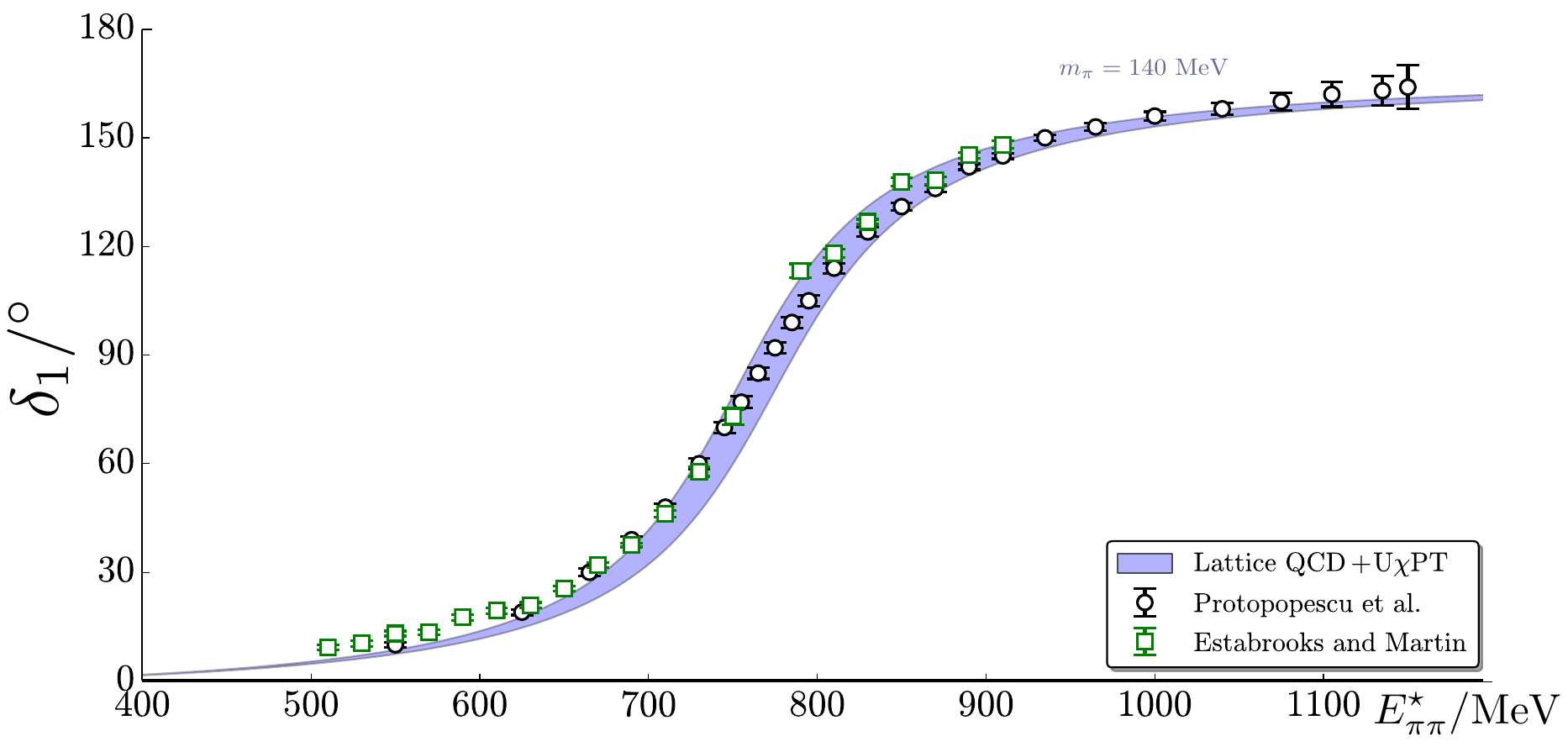}
\label{fig:UchiPT}}
\caption{
(a) Shown is the determination of the $\ell=1$ $\pi\pi$ scattering phase shift determined by the Hadron Spectrum Collaboration using two different values of the pion mass, $m_\pi\approx 240~{\rm MeV}, 400~{\rm MeV}$~\cite{Dudek:2012xn, Wilson:2015dqa}. $E^\star_{\pi\pi}$ denotes the c.m. energy of the $\pi\pi$ system. (b) Depicted by the blue band is the extrapolation performed in Ref.~\cite{Bolton:2015psa} to the physical quark masses of the $\pi\pi$ scattering phase shift determined in Ref.~\cite{Wilson:2015dqa}. The extrapolated result is compared with the experimental phase shifts, depicted here in black cicles~\cite{Protopopescu:1973sh} and green squares~\cite{Estabrooks:1974vu}.  }\label{fig:global_comparison}
\end{figure}

This formalism has been extensively implemented in the literature in the study of elastic~\cite{Dudek:2012xn, Lang:2011mn, Lang:2014yfa, Feng:2010es} and inelastic~\cite{Dudek:2016cru, Wilson:2015dqa, Wilson:2014cna, Dudek:2014qha}
resonant scattering amplitudes. In Fig.~\ref{fig:860_840_scale_set} I highlight a recent calculation of the $\pi\pi$ scattering phase shift by the Hadron Spectrum Collaboration using two different values of the quark masses corresponding to $m_\pi\approx 240~{\rm MeV}, 400~{\rm MeV}$~\cite{Dudek:2012xn, Wilson:2015dqa}. Having a physical scattering amplitude, one can proceed to compare it to the experimental one. As experimentalist are constrained to use a quark masses corresponding to $m_\pi\approx 140~{\rm MeV}$, we must first extrapolate the lattice QCD results obtained using heavy quark masses to the physical ones. This was carried out in Ref.~\cite{Bolton:2015psa}. The resulting phase shift is plotted and compared with the experiment in Fig.~\ref{fig:UchiPT}. 

Having determined the resonant scattering amplitude, one can obtain the resonance's mass and width by analytically continuing the amplitude onto the complex plane and finding its pole. This procedure, depicted in Fig.~\ref{fig:flow_charts}, is the same for lattice QCD calculations as experiment. Following this procedure, Ref.~\cite{Bolton:2015psa} found the $\rho$ pole to be $E_\rho= \left[755(2)(^{20}_{02})-\frac{i}{2}\,129(3)(^{7}_{2})\right]~{\rm MeV}$. The first uncertainty corresponds to the statistical and the second is a combination of various systematic errors, which include scale setting and uncertainties in the input masses. Its mass and width are given by the real and imaginary components of the pole position: $m_\rho={\rm Re}(E_\rho)=755(2)(^{20}_{02})~{\rm MeV}$ and $\Gamma_\rho=-2{\rm Im}(E_\rho)=\,129(3)(^{2}_{7})~{\rm MeV}$.

For sometime now it has been well understood how one can study strong decays of hadronic resonances. Until recently it was not evident how to study electroweak decays of these, for example $\rho\to\pi\gamma^\star$ transitions. Presently, the only means to do this is to determine $\pi\to\pi\pi$ infinite-volume transition amplitude, $\mathcal{H}^{\rm out}$, from finite-volume matrix element, $\langle \pi\pi\big|\mathcal{J}\big|\pi\rangle_L$, of an external current $\mathcal{J}$. It is important to recognize that the $\pi$ state is composed of a single stable hadron while the $\pi\pi$ state is composed of two. With this in mind, we can frame the $\pi\to\pi\pi$  transition as a subset of a general class of $1\to2$ transitions, where the numbers denote the number of QCD-stable hadrons present in the initial and final state. Since the $\rho$ is unstable and couples to $\pi\pi$, it must be labeled as a state with ``$2$" hadrons. 

One can find a model-independent relation between finite-volume matrix elements and infinite-volume transitions amplitudes~\cite{Briceno:2015csa, Briceno:2014uqa}
\begin{align}
\label{eq:LL}
\left| 
\langle {2}\big|\mathcal{J}\big|{1}\rangle_L\right|
=\sqrt{\frac{1}{2E_1}}
\sqrt{\mathcal{H}^{\rm in}~\mathcal{R}~\mathcal{H}^{\rm out}},
\end{align}
where $\mathcal{R}$ is a known function that depends on the spectrum, volume and the scattering amplitude. This equation is the generalization of Lellouch's and L\"uscher's original relation~\cite{Lellouch:2000pv} and it holds for any local current. In this equation $\mathcal{R}$ is a matrix in the space of partial waves and open channels, while $\mathcal{H}^{\rm in}$ and $\mathcal{H}^{\rm out}$ are row and column vectors in this space. Given the finite-volume spectrum, the $2\to2$ scattering amplitude extracted from it, and the matrix elements calculated in a finite volume, one can constrain infinite-volume transition amplitudes. This procedure is qualitatively depicted by the three blue arrows in the bottom panel of Fig.~\ref{fig:LQCD_flow}. This parallels the amplitude analysis needed to obtain transition amplitudes from experimental data for electro/photo-production processes. In the following section, I describe in further details of the implementation of this formalism by considering a specific example: $\pi^+\gamma^\star\to\pi^+\pi^0$.

\section{$\pi^+\gamma^\star\to\pi^+\pi^0$ and the unstable $\rho\to\pi\gamma^\star$  form-factor \label{sec:pigamma_to_pipi}}
The formalism presented in the previous section, Eq.~\ref{eq:LL}, holds for generic $1\to 2$ processes. Eventually it would be desirable to perform calculations of transition amplitude involving baryonic resonances, but for now processes involving mesonic degrees of freedom are simpler and serve as a natural testing ground for more these complex systems. In Sec.~\ref{sec:outlook} I review some of the challenges for calculations involving baryons. 

As a stepping stone towards the study of $N\to N^\star$ transitions, the Hadron Spectrum Collaboration recently performed an exploratory calculation of the $\pi^+\gamma^\star\to\pi^+\pi^0$ transition amplitude~\cite{Briceno:2015dca, Briceno:2016kkp}. This was done, following the prescription outlined in the previous section. First, the finite-volume matrix element, ${\langle \pi;L|{\mathcal{J}}^{\mu}_{x=0}| \pi\pi;L\rangle}$, are calculated for a range of  discrete $\pi\pi$ center of mass (c.m.) energies, $E^\star_{\pi\pi}$, and a virtuality of the photon, $Q^2=-(P_{\pi\pi}-P_{\pi})^2$. ${\mathcal{J}}^{\mu}=(2\bar{u}\gamma^\mu u-\bar{d}\gamma^\mu d)/{3}$ is the electromagnetic current and $u$ and $d$ are the up- and down-quark fields. This exploratory calculation was performed using quark masses corresponding to $m_\pi\approx400$~MeV and it was made possible by the technology developed and tested in Ref.~\cite{Shultz:2015pfa}. Having determined the matrix elements and the scattering amplitude, one can then use Eq.~\ref{eq:LL} to obtain the infinite-volume transition amplitude. 

\begin{figure}[t]
\begin{center}
\hspace*{-.7cm}                                                           
\subfigure[]{\includegraphics[scale=0.3]{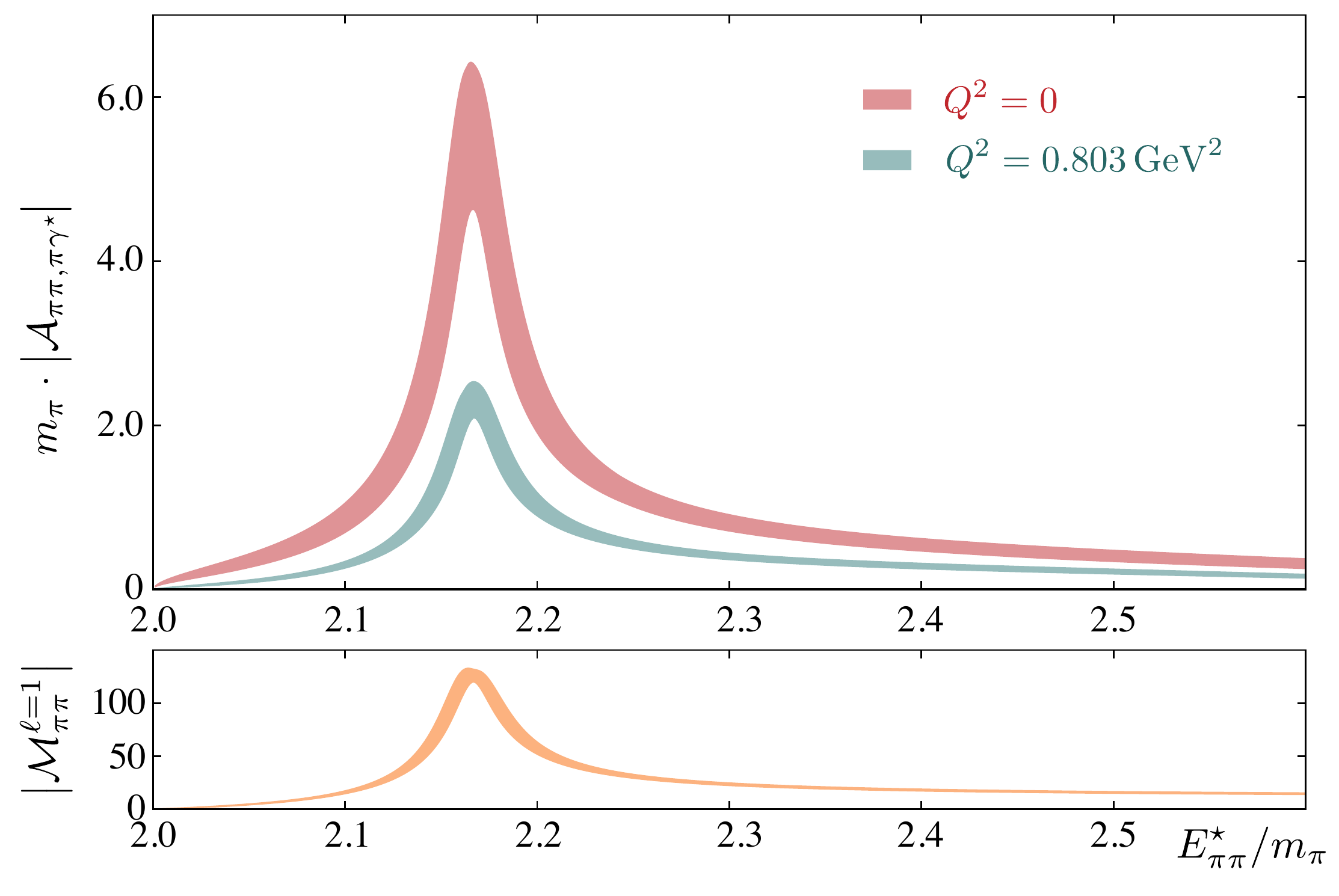}
\label{fig:slices_in_Q2}}
\hspace{1cm}
\subfigure[]{\includegraphics[scale=0.3]{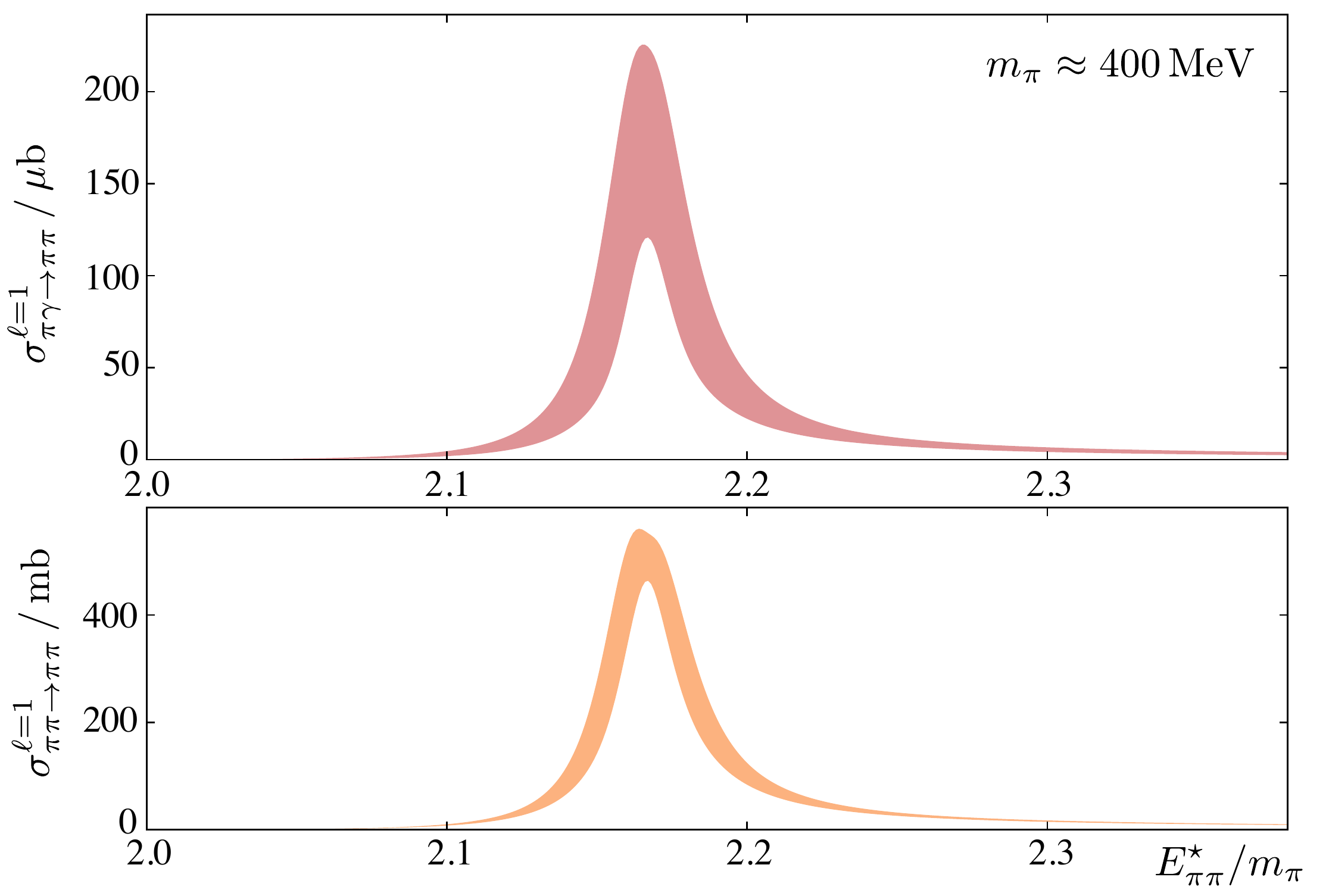}
\label{fig:cross_sec}}

\caption{ (a) Above is shown the absolute value of the transition amplitude $ {\mathcal{A}}_{\pi\pi,\pi\gamma^\star}$ as a function of the $E_{\pi\pi}^\star$ for two different values of $Q^2$ from Ref.~\cite{Briceno:2015dca, Briceno:2016kkp}. This is compared with the elastic $\ell=1$ $\pi\pi$ amplitude shown below~\cite{Dudek:2012xn}. (b) Above is shown the $\pi^+\gamma\to\pi^+\pi^0$ cross section, and it is compared to the elastic $\ell=1$ scattering cross section shown below. }
\label{fig:amplitude}
\vspace*{-.6cm}
\end{center}
\end{figure}

Given that the amplitude, $\mathcal{H}_{\pi\pi,\pi\gamma^\star}^{\mu}$, is a Lorentz vector, it is convenient to define a Lorentz scalar amplitude, ${\mathcal{A}}_{\pi\pi,\pi\gamma^\star}$. One choice to parametrize the amplitude is 
\begin{align}
\label{eq:Apipipi}
\mathcal{H}_{\pi\pi,\pi\gamma^\star}^{\mu}=
\epsilon^{\mu\nu\alpha\beta}P_{\pi,\nu}P_{\pi\pi,\alpha}\epsilon_{\beta}(\lambda_{\pi\pi},\textbf{P}_{\pi\pi})\frac{2~{\mathcal{A}}_{\pi\pi,\pi\gamma^\star}}{m_{\pi}},
\end{align}
where $\epsilon_{\beta}$ is the polarization of the $\pi\pi$ channel and $\lambda_{\pi\pi}$ is its helicity. The calculated transition amplitude is shown in Fig.~\ref{fig:slices_in_Q2} for two values of the virtuality of the photon. This amplitude is compared to the elastic scattering amplitude. One important observation is that these amplitudes exhibit the same resonant behavior. 

Another convenient illustration of the result is to plot the $\pi^+\gamma\to\pi^+\pi^0$ cross section for a real photon, $\sigma(\pi^+\gamma\to\pi^+\pi^0)$. In Fig.~\ref{fig:cross_sec}, this cross section is plotted in physical units and is contrasted with the elastic cross section. One observes that, although the resonant behavior of these two are quite similar, the latter is nearly three orders of magnitude larger. This is to be expected, as the $\pi^+\gamma\to\pi^+\pi^0$ cross section is proportional to the electromagnetic fine structure constant. 

Having constrained the transition amplitude, one can proceed to determine the $\rho\to\pi\gamma^\star$ form factor. This can only be rigorously defined as the residue of the transition amplitude at the $\rho$-pole. Given that the amplitude is only constrained on the real axis, one can expect the form factor to be dependent upon the parametrization chosen for the $E^\star_{\pi\pi}$ dependence. By trying a range of choices, a systematic uncertainty can be estimated, which we find to be small for the narrow $\rho$ resonance. This is the non-trivial step depicted by the magenta arrows in the lower panel of Fig.~\ref{fig:LQCD_flow}. The dashed lines connecting the partial wave amplitude emphasize that one may use information obtained from the partial wave amplitude in order to perform the analytic continuation. This is not a necessity, since both the transition amplitudes and elastic scattering amplitudes should have the same pole structure.

\begin{figure}[t]
\begin{center}
{\includegraphics[scale=0.4]{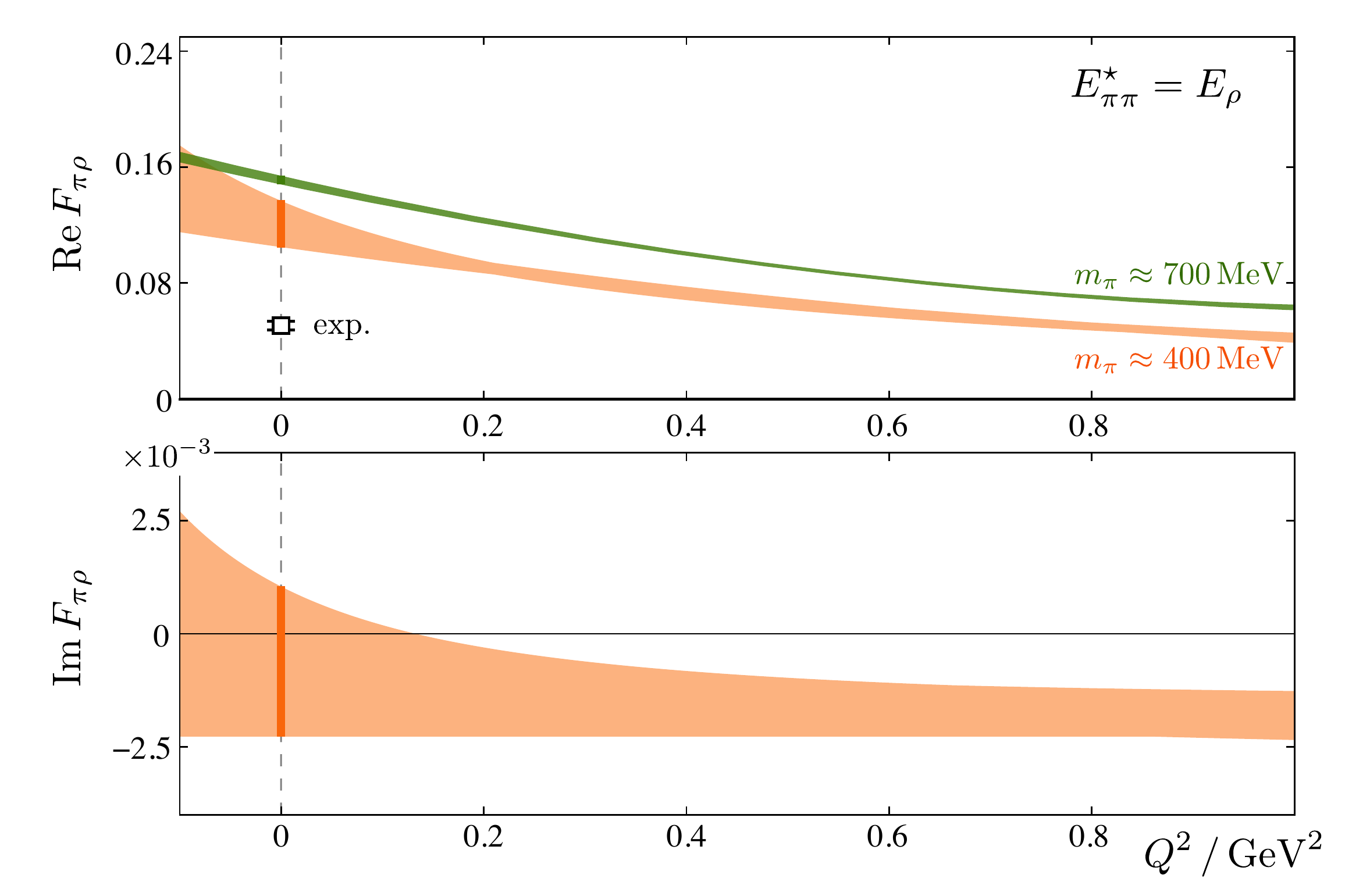}}
\caption{
Above is shown the real component of the $\rho\to\pi\gamma^\star$ form factor determined in using $m_\pi\approx 400$~MeV~\cite{Briceno:2015dca, Briceno:2016kkp} [orange band]. This is evaluated at the $\rho$ pole. For comparison are shown the form factor determined in Ref.~\cite{Shultz:2015pfa} [green band] using $m_\pi\approx 700$~MeV, where the $\rho$ resonance is QCD stable, and the experimentally determined $\rho\pi$ photocoupling~\cite{Huston:1986wi, Capraro:1987rp}. The lower panel shows the imaginary component of the form factor obtained for $m_\pi\approx 400$~MeV.
}
\label{fig:Fpirho_plot}
\vspace*{-.6cm}
\end{center}
\end{figure}

The definition used for the form factor, $F(\Epipi^\star,Q^2)$, is the following,
\begin{align}
\label{eq:AtoF}
{\mathcal{A}}_{\pi\pi,\pi\gamma^\star}
&={\F(\Epipi^\star,Q^2)} \sin\delta_1~\sqrt{\frac{16\pi }{q^\star_{\pi\pi}\Gamma_{P}(\Epipi^\star)}}~e^{i\delta_1},
\end{align}
where $\Gamma_{P}(\Epipi^\star)$ is the energy-dependent width of the $\rho$ which can be constrained from the elastic $\pi\pi$ scattering amplitude. Using this relation, Hadron Spectrum Collaboration found the $\rho\to\pi\gamma^\star$ form factor show in Fig.~\ref{fig:Fpirho_plot}. This was done using two values of the quark masses. The first corresponds to $m_\pi\approx 700$~MeV~\cite{Shultz:2015pfa} where the $\rho$ is stable, and the second for $m_\pi\approx 400$ where the $\rho$ decays to $\pi\pi$~\cite{Briceno:2015dca, Briceno:2016kkp}. Since the $\rho$ is stable for the  $m_\pi\approx 700$~MeV calculation, the finite formalism was not necessary in its analysis.

One can observe two important facts concerning the behavior of the form factor as a function of the quark masses. First, it appears to follow a natural trend towards the physical point, as the $m_\pi\approx 400$~MeV ensemble is closer to the experimental point than the $m_\pi\approx 700$~MeV ensemble. The second important observation is that the $m_\pi\approx 400$~MeV  $\rho$-pole is located off the Re[s]-axis. As a result, the form factor, as defined in Eq.~\ref{eq:AtoF}, can in general be complex. The imaginary component of the form factor is found to be approximately two orders of magnitude smaller than its real component. This can be understood by the fact that the $\rho$ is rather narrow for these quark masses.

\section{Outlook to $N\to N^\star$ transitions from QCD \label{sec:outlook}}

Having performed the first calculation of a resonant radiative transition process, one could naturally ask ``\emph{why not perform a similar calculation for $N\gamma^\star\to N^\star$?}". 
\footnote{For a complimentary discussion on the present status of $N^\star$ studies from lattice QCD, I point the reader to Ref.~\cite{Richards:2016hki}.}
Here I briefly discuss the obtasbles associated with these calculations and give an outlook for the future. There are several challenges, some of which I list: more open thresholds, three-particle thresholds might be important, larger number of contractions, deterioration of signal, larger number of partial waves, and partial-wave mixing. To clarify what is meant by each one of these, I describe them below.

The first two of these point are important since the $N\gamma^\star\to N^\star$ form factor is only determinable from the residue of the transition amplitude for $N\gamma^\star\to N\pi,N\pi\pi,N\eta,\ldots$. The formalism presented in Ref.~\cite{Briceno:2015csa, Briceno:2014uqa} accommodates energies above any number of two-body thresholds but below three-particle thresholds. This is of particular importance for resonances like the \emph{Roper} $N(1440)$, which experimentally couples approximately $30-40\%$ of the time to $N\pi\pi$. A universal formalism for three-particles in a finite volume is not yet in place, but important steps have been taken in this direction~\cite{Hansen:2014eka,  Hansen:2015zga, Polejaeva:2012ut, Briceno:2012rv}. 

Given that baryonic systems have a larger number of valence quarks, when computing correlations functions this leads to a larger number of Wick contractions than in the mesonic sector. This is a challenge but certainly not a limitation. For instance, Ref.~\cite{Lang:2012db} has already presented an exploratory calculation of resonant $N\pi$ scattering in the negative parity sector, and recently Ref.~\cite{Detmold:2015qwf} presented a calculation of meson-baryon phase-shifts in various non-resonant channels. As is clear from these and all other studies in the baryonic sector, the signal deteriorates quicker for baryonic systems than mesonic ones. Reducing the statistical noise of these calculations is possible by lengthening their computational running time.

Lastly, given that baryons carry nonzero intrinsic spin, there is a larger number of partial waves for baryonic systems than in systems of spin-0. Due to the reduction of rotational symmetry these mix, as is dictated by Eqs.~\ref{eq:QC_master} and \ref{eq:LL}. Unlike $\pi\pi$, for the $N\pi$ system positive- and negative-parity partial waves will mix when the system is boosted. This would imply that, for example, in a rigorous lattice QCD calculation one would have to simultaneously study the $N(1440)$, $N(1520)$, $N(1535)$, $\ldots$ resonances. This is a challenge, but it has been previously addressed in, for example, the $\pi K,\eta K$ channels by the Hadron Spectrum Collaboration~\cite{Wilson:2014cna, Dudek:2014qha}. In this work, the authors performed a simultaneously determination of the $\kappa$, $K^\star_0$, $K^\star_1,$ and $K^\star_2$ resonance poles.

In summary, the historical limitations associated with the study of resonant hadrons are currently being overcome. There is still more development needed, but this is currently being addressed by the lattice QCD community. Therefore, it is not unrealistic to expect calculation of resonant elastic/inelastic scattering and transitions processes in the baryonic sector in the near future. As in the mesonic sector, the first calculations will be of scattering processes, followed by transition processes involving electroweak external currents.

\section{Final Remarks \label{sec:final_remarks}}
Lattice QCD has proven to be a remarkable tool for the determination of low-lying, QCD-stable states. The determination of properties of resonances via lattice QCD has been historically limited. In the past few years we have witnessed a tremendous amount of progress that has opened the doors towards increasingly complex and important observables. It is worth reiterating that we are entering an exciting time for the study of few-body systems, and we should expect many more studies of phenomenologically interesting processes directly from QCD. In this talk I have highlighted some important developments towards this goal. 

\begin{acknowledgements}

 I would like to thank my colleagues within the Hadron Spectrum Collaboration, in particular Jozef Dudek, Robert Edwards, Christian Shultz, Christopher Thomas and David Wilson, for granting me permission to share the results of their hard work. Also, I would like to thank my other collaborators, Daniel Bolton and Maxwell Hansen, whose work I have also presented in this talk. 

 \end{acknowledgements}


\bibliographystyle{apsrev} 
\bibliography{bibi} 

\end{document}